\documentclass{article}
\usepackage{spconf}

\usepackage{amsfonts,url,graphicx}
\usepackage[cmex10]{amsmath}

\usepackage[tight,footnotesize]{subfigure}

\usepackage{slashbox}

\newcommand{\cv}[1]{\mathbf{#1}}

\newcommand{\tr}{^{\rm T}}

\newcommand{\her}{^{\rm H}}

\newcommand{\Id}{{\bf I}}
\newcommand{\Od}{{\bf O}}

\newcommand{\NT}{N_{\rm T}}
\newcommand{\NR}{N_{\rm R}}

\renewcommand{\v}[1]{\underline{\mathbf{#1}}}

\title{On the Achievability of Interference Alignment in the \emph{K}-User Constant MIMO Interference Channel}

\name{Roland Tresch, Maxime Guillaud, and Erwin Riegler\thanks{This work was supported by the STREP project IST-026905 (MASCOT) within the sixth framework programme of the European Commission, and by the Comet competence center program of the Austrian government through the I0 project of the Vienna Telecommunications Research Center (ftw.).}}
\address{FTW (Telecommunications Research Center Vienna)\\
         Vienna, Austria\\
         e-mail: \tt{\{tresch,guillaud,riegler\}@ftw.at}}

\begin{document}

\ninept


%


\maketitle

\begin{abstract}
Interference alignment in the \emph{K}-user MIMO interference channel with constant channel coefficients is considered. A novel constructive method for finding the interference alignment solution is proposed for the case where the number of transmit antennas equals the number of receive antennas ($\NT=\NR=N$), the number of transmitter-receiver pairs equals $K=N+1$, and all interference alignment multiplexing gains are one. The core of the method consists of solving an eigenvalue problem that incorporates the channel matrices of all interfering links. This procedure provides insight into the feasibility of signal vector spaces alignment schemes in finite dimensional MIMO interference channels.
\end{abstract}



%

\section{Introduction}
\label{section_introduction}

Interference alignment was first considered in \cite{maddahali_isit06} as a coding technique for the two-user Multiple-Input Multiple-Output (MIMO) X channel, where it was shown to achieve multiplexing gains strictly higher than that of the embedded MIMO interference channel (IC), multiple-access channel (MAC), and broadcast channel (BC) taken separately. While requiring perfect channel knowledge, this coding technique is based only on linear precoding at the transmitters and zero-forcing at the receivers. The degrees of freedom region for the X channel was analyzed in \cite{Jafarit08} for an arbitrary number of antennas per user. This transmission technique was later generalized to the $K$-user interference channel \cite{Cadambeit08}, where it was shown to achieve almost surely a sum-rate multiplexing gain of $\frac{K}{2}$ per time, frequency and antenna dimension. In comparison, independent operation of $K$  \emph{isolated} point-to-point links would incur a sum-rate multiplexing gain of $K$ per dimension. This indicates that interference alignment allows virtually interference-free communications at the cost of each user exploiting only half of the available degrees of freedom. Thanks to the alignment of all interfering signals in the same subspace from the point of view of all receivers simultaneously, interference can be removed simply through zero-forcing filtering.

In \cite{Cadambeit08}, an explicit formulation of the precoding vectors achieving interference alignment is presented for single-antenna nodes with time-varying channels. In the multiple-antenna case, no such closed-form solution is known, although achievability results on multiplexing gains are available. In \cite{Gou08}, an inner bound and an outer bound on the total number of degrees of freedom for the MIMO case are presented. Achievability results on the total number of achievable degrees of freedom are also presented in \cite{cadambe-2009}. An iterative algorithm was introduced in \cite{Gomadamit08} to find numerically the precoding matrices achieving interference alignment. Little results have been obtained regarding \emph{constant} MIMO interference channels. Some results concerning the $K$=3 user case are presented in \cite{Cadambeit08}.\\

The goal of this contribution is to give a constructive proof of the achievability of interference alignment in the $K$-user constant MIMO interference channels, for certain dimensions of the channel matrices. The proof in Section \ref{section_Novel_Alignment_Method} deals with the case where the number of transmit antennas equals the number of receive antennas $(\NT=\NR=N)$, the number of transmitter-receiver pairs equals $K=N+1$. For those channel dimensions, if the channel coefficients are drawn independently from a continuous distribution, we show that one degree of freedom is achievable per user with probability 1. The method provides insight into the general $K$-user MIMO interference channel.\\

This article is organized as follows: the system model, interference alignment, and the iterative algorithm are introduced in Section \ref{section_sysmodel}. The constructive alignment method is introduced in Section \ref{section_Novel_Alignment_Method}. Section \ref{section_Unfeasible_Network_Setting} describes an interference channel setting where interference alignment is unfeasible and Section \ref{section_Conjecture} gives further insights obtained by the proposed method.

\section{System Model}
\label{section_sysmodel}
In this section, we first introduce the system model for arbitrary (non-square) channel dimensions, and introduce the concept of interference alignment.
We consider a $K$-user MIMO interference channel with transmitters equipped with $\NT$ antennas and receivers with $\NR$ antennas. The MIMO channel is assumed to be frequency-flat and the channel coefficients are drawn independently from a continuous distribution.

We focus our attention on the $i^{\textrm{th}}$ receiver $(1\leq i \leq K)$, which receives interference from other transmitters $j\neq i$ in addition to its intended signal.
The discrete-time channel model at a given time instance is given by
\begin{align}\label{equ_yibar}
\v{y}_i = & \cv{H}_{ii}\cv{V}_{i} \v{s}_i +
\sum_{j=1,j\neq i}^{K}\cv{H}_{ij} \cv{V}_{j} \v{s}_j + \v{n}_i,
\end{align}
where $\v{s}_i \in \mathbb{C}^{d_i\times 1}$ is a vector representing the signals from transmitter $i$. $\cv{V}_{i}\in \mathbb{C}^{\NT\times d_i}$ is the precoding matrix at transmitter $i$. $\cv{H}_{ii}\in \mathbb{C}^{\NR\times \NT}$ and $\cv{H}_{ij}\in \mathbb{C}^{\NR\times \NT}$ are $\NR\times \NT$ matrices representing the MIMO channels of the link between intended communication pairs and the interfering link between transmitter $j$ and unintended receiver $i$, respectively, experienced by user $i$. $\v{n}_i$ accounts for the thermal noise generated in the radio frequency front-end of the receiver and interference from sources other than the interfering transmitters.\\

Interference alignment (IA) is achieved with degrees of freedom $(d_1,\ldots d_K)$ (each $d_i$ corresponds to the multiplexing gain achieved for a transmitter-receiver pair, i.e. $d_i$ streams per transmitter are spatially pre-coded at transmitter $i$) iff there exists $\NT\times d_i$ truncated unitary matrices (precoding matrices) $\cv{V}_{i}$ and $\NR\times d_i$ truncated unitary matrices (zero-forcing interference suppression matrices) $\cv{U}_{i}$ such that, for $i=1,\hdots,K$,
\begin{align}
\cv{U}_{i}\her \cv{H}_{ij}\cv{V}_{j} &= 0,\forall j\neq i, \quad \mathrm{and} \label{equ_IAcondA}\\
\textrm{rank}\left(\cv{U}_{i}\her \cv{H}_{ii}\cv{V}_{i} \right)&= d_i. \label{equ_IAcondB}
\end{align}
An iterative algorithm \cite[Algorithm 1]{Gomadamit08}, which is based on the minimization of an interference leakage metric (zero leakage is equivalent to the system of equations in \eqref{equ_IAcondA} and \eqref{equ_IAcondB}), was introduced to find the precoding matrices and to verify the achievability of interference channel settings. At every iteration, the algorithm from \cite{Gomadamit08} involves the computation of $K$ eigenvalue problems. Furthermore, depending on the interference channel setting the convergence speed can vary significantly. Nevertheless, the iterative algorithm provides numerical insights into the feasibility of IA for the $K$-user $\NT\times \NR$ MIMO interference channel with any IA multiplexing gains.

If a solution to the system \eqref{equ_IAcondA} and \eqref{equ_IAcondB} is found, the received signal of user $i$ after interference suppression yields
\begin{align}\label{equ_yibar_zf}
\bar{\v{y}}_i &= \cv{U}_{i}\her\cv{H}_{ii}\cv{V}_{i} \v{s}_i + \sum_{j\neq i}\cv{U}_{i}\her\cv{H}_{ij} \cv{V}_{j} \v{s}_j + \cv{U}_{i}\her\v{n}_i,\nonumber \\
              &= \cv{U}_{i}\her\cv{H}_{ii}\cv{V}_{i} \v{s}_i+ \bar{\v{n}}_i,
\end{align}
where we used the fact that the interference term in \eqref{equ_yibar} is perfectly suppressed, due to \eqref{equ_IAcondA}. Note that the energy of the signal part that lies in the interference subspace is lost. However, this power loss does not matter in the context of the degrees of freedom analysis. 

\section{Constructive Alignment Method}
\label{section_Novel_Alignment_Method}

In this section, we focus on the case where the number of transmit antennas equals the number of receive antenna ($\NT=\NR=N$) and the number of transmitter-receiver pairs $K=N+1$. Furthermore, and conversely to the achievability schemes of \cite{cadambe-2009} where only the total number of degrees of freedom is considered, we assume that each user gets one interference-free degree of freedom, i.e., $d_i=1\,\forall i$. Throughout the paper, if $d_i=1\,\forall i$, we will denote the $\cv{V}_i$'s as the $N\times 1$ unit norm precoding vectors $\v{v}_i \,\forall i$ and the $\cv{U}_i$'s as the $N\times 1$ unit norm interference suppression vectors $\v{u}_i\,\forall i$.
\begin{figure}[h]
  \centering
  \includegraphics[width=7cm]{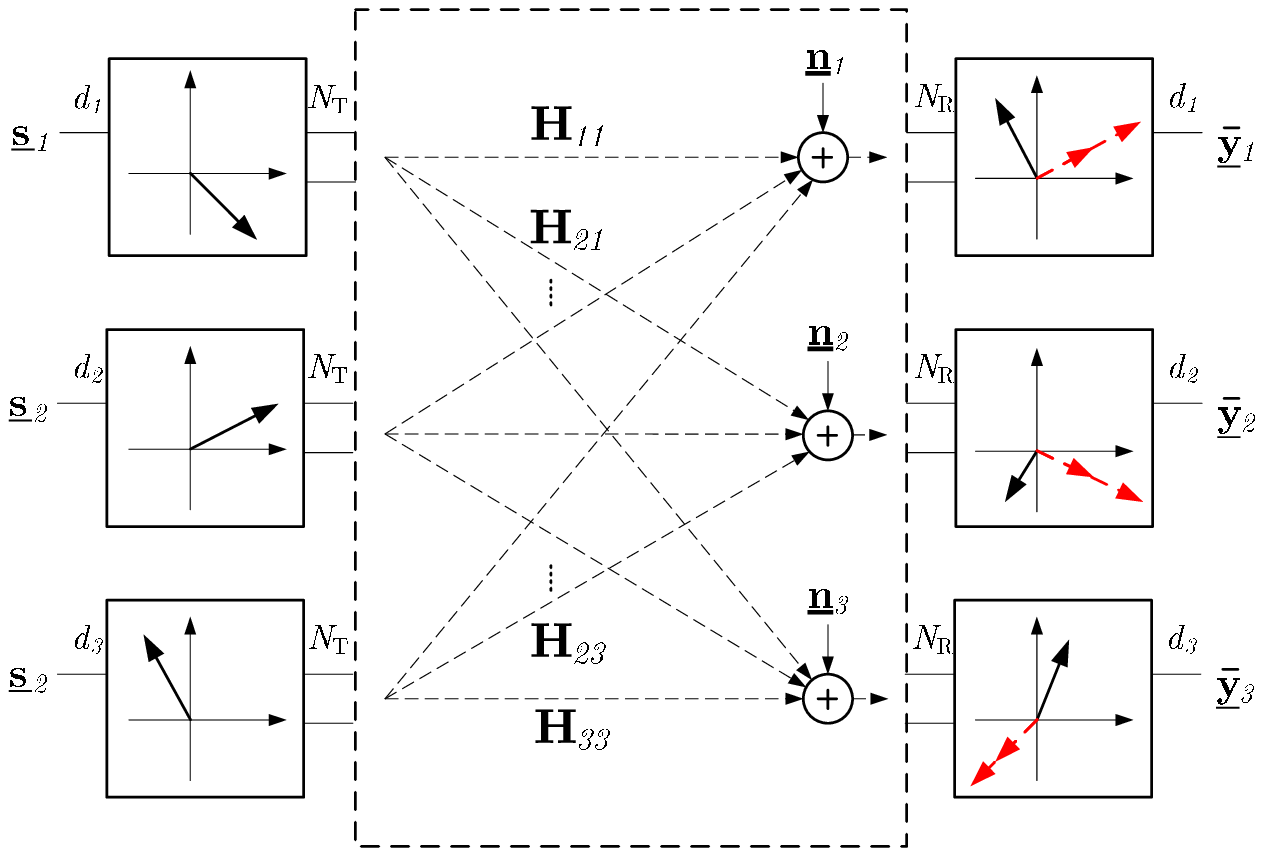}
  \caption{IA incorporating three transmitter-receiver pairs.}
  \label{Fig:systemmodelK3_2x2}
\end{figure}
Fig. \ref{Fig:systemmodelK3_2x2} depicts the IA solution for a $2\times 2$ MIMO interference channel that incorporates three transmitter-receiver pairs. The precoding leads to the following situation in the $N$-dimensional receive signal space at receiver $i$: the interference from transmitter $j$ and $k$, $(j,k)\neq i$, aligns perfectly in a one-dimensional subspace. We observe that every transmitter has to sacrifice in the number of signal streams compared to the case where only a single transmitter-receiver pair is present, where $\min(\NR,\NT)=2$ streams could be transmitted over the eigenmodes of the MIMO channel.

In the following, we describe a novel alignment method that provides a solution to the IA conditions \eqref{equ_IAcondA} and \eqref{equ_IAcondB} by reformulating \eqref{equ_IAcondA} as an eigenvalue problem.

\subsection{Equivalent Eigenvalue Problem}
\label{subsection_Equivalent_Eigenvalue_Problem}

Let us consider the $N$-dimensional receive signal spaces. Each $\v{u}_{i}$ in the IA condition \eqref{equ_IAcondA} defines a subspace of codimension one. Furthermore, \eqref{equ_IAcondA} is equivalent to requiring that the interfering signals $\cv{H}_{ij}\v{v}_{j}\,\forall j\neq i$ at each receiver lie in subspaces of dimension at most $N-1$. Thus, since $K=N+1$, this is equivalent to linear dependency of the $K-1=N$ interfering signals at each receiver. That is, there exist $\mu_{ij}$'s, $j\neq i$, where at least one of the $\mu_{ij}$'s is nonzero, such that
\begin{align}\label{equ_cond_K_Np1}
   \sum_{j=1,j\neq i}^K\mu_{ij} \cv{H}_{ij}\v{v}_{j} = \underline{0}_{N},
\end{align}
with $\underline{0}_{N}$ the $N \times 1$ all zero vector.

From now on, we tackle \eqref{equ_IAcondA} simultaneously for all receivers $i$. Therefore, we stack the precoding vectors in $\v{v}=[\v{v}_{1}\tr \, \v{v}_{2}\tr \, \ldots \, \v{v}_{K}\tr]\tr$  and define
\begin{align}\label{equ_H}
\small
\cv{H}=
\left[\begin{array}{cccc}
\Od_{N} & \mu_{12} \cv{H}_{12} & \ldots & \mu_{1K} \cv{H}_{1K} \\
\mu_{21} \cv{H}_{21} & \Od_{N} & \ldots & \vdots \\
\vdots & \ddots & \ldots &  \mu_{(K-1)K} \cv{H}_{(K-1)K} \\
\mu_{K1} \cv{H}_{K1} & \ldots & \ldots & \Od_{N}
\end{array}\right],
\end{align}
where $\Od_{N}$ is the $N \times N$ all zero matrix. $\cv{H}$ incorporates the channel matrices of all interfering links in the interference channel. Using this notation, \eqref{equ_IAcondA} is equivalent to
\begin{align}
\cv{H}\cdot \v{v} &= \underline{0}_{KN},\label{equ_condHv0}
\end{align}
for any given $\mu_{ij}$'s defined by \eqref{equ_cond_K_Np1}, with $\underline{0}_{KN}$ the $KN \times 1$ all zero vector. \eqref{equ_condHv0} is invariant under left multiplication with a matrix $\cv{P}$ that shifts rows. Thus, we get a nonzero block diagonal $\cv{B}=\textrm{diag}\{\mu_{K1}$ $\cv{H}_{K1} \, \ldots \, \mu_{(K-1)K} \cv{H}_{(K-1)K}\}$ in $\cv{P}\cv{H}$, with the permutation matrix
\begin{align}\label{equ_P}
\small
\cv{P}=
\left[\begin{array}{cccc}
\Od_{N} & \ldots & \Od_{N} & \Id_{N} \\
\Id_{N} & \Od_{N} & \ddots & \Od_{N} \\
\vdots & \ddots & \ddots & \vdots \\
\Od_{N} & \ldots & \Id_{N} & \Od_{N}
\end{array}\right].
\end{align}
Furthermore, since all channel matrices are of rank $N$ with probability 1 and if we assume that $(\mu_{K1},\ldots,\mu_{(K-1)K})\neq 0$, we can pre-multiply the permuted matrix $\cv{P}\cv{H}$ with the inverse of the block diagonal and define a matrix
\begin{align}\label{equ_Hcomp}
\widehat{\cv{H}} &= -\lambda(\cv{B}^{-1}\cv{P}\cv{H}-\Id_{KN}),
\end{align}
with the free parameter $\lambda\neq 0$, which turns \eqref{equ_condHv0} into the standard eigenvalue equation
\begin{align}\label{equ_condHcomp}
\left(\widehat{\cv{H}}-\lambda\Id_{KN} \right)\cdot \v{v}=\underline{0}_{KN}.
\end{align}
It is clear from \eqref{equ_condHv0}-\eqref{equ_condHcomp} that any $\v{v}$ associated to a nonzero eigenvalue of $\widehat{\cv{H}}$ fulfills \eqref{equ_condHv0}, for any choice of the $(\mu_{K1},\ldots,\mu_{(K-1)K})$ for which $\cv{B}$ is not singular.
That is, if we define $\nu_{ij}=-(\mu_{lj}/\mu_{li})\lambda$ with $l=\textrm{mod}(i-2,K)+1$, it is sufficient to choose the $\nu_{ij}$'s such that $\widehat{\cv{H}}$ has at least one non-zero eigenvalue, to solve the interference alignment problem. Note that due to the structure of $\widehat{\cv{H}}$ (depicted in eq. \eqref{equ_Hcomp_ex}, next page), choosing e.g. $\nu_{ij}=1, \ \forall (i,j)$ guarantees that this happens with probability 1 when the channel coefficients are drawn independently from a continuous distribution.

\begin{figure*}[!t]
\small
\begin{align}\label{equ_Hcomp_ex}
\widehat{\cv{H}}=
\left[\begin{array}{ccccc}
\Od_{N} & \nu_{12}\cv{H}_{K1}^{-1}\cv{H}_{K2} & \ldots & \ldots & \Od_{N} \\
\Od_{N} & \Od_{N} & \nu_{23}\cv{H}_{12}^{-1}\cv{H}_{13} & \ldots & \nu_{K2}\cv{H}_{12}^{-1}\cv{H}_{1K} \\
\nu_{31}\cv{H}_{23}^{-1}\cv{H}_{21} & \Od_{N} & \ddots  & \ldots & \vdots \\
\vdots & \ddots & \ddots &  \ldots &  \nu_{(K-1)K}\cv{H}_{(K-2)(K-1)}^{-1}\cv{H}_{(K-2)K} \\
\nu_{K1}\cv{H}_{(K-1)K}^{-1}\cv{H}_{(K-1)1} & \ldots & \nu_{K(K-2)}\cv{H}_{(K-1)K}^{-1}\cv{H}_{(K-1)(K-2)} & \ldots & \Od_{N}
\end{array}\right].
\end{align}
\hrulefill
\vspace*{4pt}
\end{figure*}
Finally, in order to provide the complete solution to \eqref{equ_IAcondA}, the  vectors $\v{v}_{1},\,\v{v}_{2},\,\ldots,\, \v{v}_{K}$ must be normalized, and the $\v{u}_{i}$ can be determined by finding any unit norm vector in the orthogonal complement of the subspace (of dimension at most $N-1$) spanned by interfering signals $\cv{H}_{ij}\v{v}_{j}\,\forall j\neq i$ at each receiver $i$.

Note that the above technique achieves a total of $K$ degrees of freedom over the $K$-user constant interference channel with $N\times N$ matrices. While for $K=3$ some methods achieving more degrees of freedom are known \cite{Cadambeit08}, we are not aware of similar constructive results for the case $K>3$.

\subsection{Three User Network}
\label{subsection_Three_User_Network}

In order to illustrate the proposed method, we will again consider the $K=3$ user $2\times 2$ MIMO interference channel with IA multiplexing gains $(1,1,1)$ as depicted in Fig. \ref{Fig:systemmodelK3_2x2}. Deriving the equivalent eigenvalue problem \eqref{equ_condHcomp} leads to further insight that we will deliver in the following.

Following the method introduced in Subsection \ref{subsection_Equivalent_Eigenvalue_Problem}, \eqref{equ_condHcomp} leads to the sufficient condition for an IA solution, i.e.,
\begin{align}\label{equ_evK3_2x2}
\small
& \Bigg(\underbrace{\left[\begin{array}{ccc}
\Od_{N} & \nu_{12} \cv{H}_{31}^{-1}\cv{H}_{32} & \Od_{N} \\
\Od_{N} & \Od_{N} &  \nu_{23} \cv{H}_{12}^{-1}\cv{H}_{13} \\
\nu_{31} \cv{H}_{23}^{-1}\cv{H}_{21} & \Od_{N} & \Od_{N}
\end{array}\right]}_{\widehat{\cv{H}}}-\lambda \Id_{KN} \Bigg)
\cdot \v{v} \nonumber\\ &= \underline{0}_{KN},
\end{align}
with any $\nu_{12}=-(\mu_{32}/\mu_{31})\lambda$, $\nu_{23}=-(\mu_{13}/\mu_{12})\lambda$ and
$\nu_{31}=-(\mu_{21}/\mu_{23})\lambda$. Clearly, the eigenvalue problem \eqref{equ_evK3_2x2} can be reformulated in terms of the characteristic equation as follows:
\begin{align}
\det\left(\widehat{\cv{H}}-\lambda\Id_{KN}\right)=0\label{equ_detK3_2x2}.
\end{align}
Interestingly, in the $K=3$ user $2\times 2$ MIMO interference channel, we can further exploit the structure of $\widehat{\cv{H}}$ and rewrite \eqref{equ_detK3_2x2} as
\begin{align}\label{equ_detK3_2x2_2}
\det\left(\nu\cv{H}_{31}^{-1}\cv{H}_{32}\cv{H}_{12}^{-1}\cv{H}_{13}\cv{H}_{23}^{-1}\cv{H}_{21} -\lambda\Id_{N}\right)&=0 \nonumber \\
\Leftrightarrow\quad \det\left(\widehat{\cv{H}}^{'} -\lambda\Id_{N}\right)&=0,
\end{align}
with $\nu=\nu_{12}\nu_{23}\nu_{31}$ and $\widehat{\cv{H}}^{'} = \nu\cv{H}_{31}^{-1}\cv{H}_{32}\cv{H}_{12}^{-1}\cv{H}_{13}\cv{H}_{23}^{-1}\cv{H}_{21}$. Here, we used the fact that $\det\left(\begin{array}{cc}
                                   A & B \\
                                   C & D
                                 \end{array}
\right) = \det(A)-\det(B-CA^{-1}B)$ and that the channel matrices are full rank with probability 1.

Eq. \eqref{equ_detK3_2x2_2} can be derived in a different way. We proceed along a shortcut in our proposed method by introducing a ``loop equation'' deduced from \eqref{equ_IAcondA}.

The interference suppression vectors $\v{u}_i$ define subspaces of codimension one in the two dimensional receive signal spaces. Therefore, the space spanned by the interfering signals $\cv{H}_{ij}\v{v}_j \,\forall j\neq i$ must be one dimensional, i.e., the interference signals are collinear. We rewrite \eqref{equ_cond_K_Np1} as
\begin{align}
\left. \begin{array}{l}
\v{u}_{1}\her \cv{H}_{12}\v{v}_{2} = 0 \\
\v{u}_{1}\her \cv{H}_{13}\v{v}_{3} = 0 \\
\end{array} \right\} \, \v{v}_{2} &= -\frac{\mu_{13}}{\mu_{12}} \cv{H}_{12}^{-1}\cv{H}_{13}\v{v}_{3}, \label{equ_condA_1}\\
\left. \begin{array}{l}
\v{u}_{2}\her \cv{H}_{21}\v{v}_{1} = 0 \\
\v{u}_{2}\her \cv{H}_{23}\v{v}_{3} = 0 \\
\end{array} \right\} \, \v{v}_{3} &= -\frac{\mu_{21}}{\mu_{23}} \cv{H}_{23}^{-1}\cv{H}_{21}\v{v}_{1}, \label{equ_condA_2}\\
\left. \begin{array}{l}
\v{u}_{3}\her \cv{H}_{31}\v{v}_{1} = 0 \\
\v{u}_{3}\her \cv{H}_{32}\v{v}_{2} = 0 \\
\end{array} \right\} \, \v{v}_{1} &= -\frac{\mu_{32}}{\mu_{31}} \cv{H}_{31}^{-1}\cv{H}_{32}\v{v}_{2} \label{equ_condA_3},
\end{align}
where we used the fact that all channel matrices are of full rank with probability 1. Inserting \eqref{equ_condA_1} and \eqref{equ_condA_2} in \eqref{equ_condA_3}, we can write
\begin{align}
\v{v}_{1} = -\frac{\mu_{32}}{\mu_{31}}\frac{\mu_{13}}{\mu_{12}}\frac{\mu_{21}}{\mu_{23}}\cv{H}_{31}^{-1}\cv{H}_{32} \cv{H}_{12}^{-1}\cv{H}_{13}\cv{H}_{23}^{-1}\cv{H}_{21}\v{v}_{1} \label{equ_condA_ev1}.
\end{align}
Thus, \eqref{equ_condA_ev1} leads to a similar eigenvalue problem like \eqref{equ_detK3_2x2_2}.
$\v{v}_{3}$ and $\v{v}_{2}$ are obtained from eqs. \eqref{equ_condA_2} and \eqref{equ_condA_1}, respectively. Again, the full solution to eq. \eqref{equ_IAcondA} is found by normalizing the precoding vectors $\v{v}_{1}, \v{v}_{2}$ and $\v{v}_{3}$, and by choosing the $\v{u}_{i}$'s in the orthogonal complement of the interference subspace, which is guaranteed to be of dimension at least 1.

Note that for even $N$, the result presented here for the three user interference channel matches that of \cite[Appendix IV]{Cadambeit08} (in terms of the achieved degrees of freedom and of the expressions of the precoding vectors), while for the case of odd $N$, the solution proposed here does not require to consider channel extensions, conversely to \cite[Appendix V]{Cadambeit08}.

\section{Unfeasible Network Setting}
\label{section_Unfeasible_Network_Setting}

So far, we have restricted our attention to the $N \times N$ MIMO interference channel with $K=N+1$ users. Let us next consider the $K=4$ user $2\times2$ interference channel, i.e., here $K=N+2$ which is the setting depicted in Fig. \ref{Fig:systemmodelK3_2x2} with one additional transmitter-receiver pair present. Note that the bounds \cite[Theorems 1 and 3]{Gou08} available for the time-varying case do not apply here, since we consider the constant channel case, and it is therefore not possible to conclude a priori about the existence of a solution.
We will now show that interference alignment with one degree of freedom per user is not achievable in this case.

Let us consider receiver $i$. The interference suppression vectors $\v{u}_i$ define subspaces of codimension one in the two dimensional receive signal spaces. Therefore, the space spanned by the interfering signals $\cv{H}_{ij}\v{v}_j\, \forall j\neq i$ must be one dimensional, i.e., here, the three interfering signals must be collinear at each receiver. Thus, there exist $(\mu_{ij}, \kappa_{ij}, \mu_{ik}, \mu_{il})\neq 0$, such that for receiver $i$: $\mu_{ij}\cv{H}_{ij}\v{v}_{j}= \mu_{ik}\cv{H}_{ik}\v{v}_{k}$ and $\kappa_{ij}\cv{H}_{ij}\v{v}_{j}= \mu_{il}\cv{H}_{il}\v{v}_{l}$. Therefore the IA conditions write\\
\begin{minipage}[h]{4.2cm}
\centering
\begin{align}
\v{v}_{2} &=\frac{\mu_{13}}{\mu_{12}} \cv{H}_{12}^{-1}\cv{H}_{13}\v{v}_{3}\label{equ_condE1},\\
\v{v}_{2} &=\frac{\mu_{14}}{\kappa_{12}} \cv{H}_{12}^{-1}\cv{H}_{14}\v{v}_{4}\label{equ_condE2},
\end{align}
\begin{align}
\v{v}_{1} &=\frac{\mu_{23}}{\mu_{21}} \cv{H}_{21}^{-1}\cv{H}_{23}\v{v}_{3}\label{equ_condE3},\\
\v{v}_{1} &=\frac{\mu_{24}}{\kappa_{21}} \cv{H}_{21}^{-1}\cv{H}_{24}\v{v}_{4}\label{equ_condE4},
\end{align}
\end{minipage}
\begin{minipage}[h]{4.2cm}
\centering
\begin{align}
\v{v}_{4} &=\frac{\mu_{31}}{\mu_{34}} \cv{H}_{34}^{-1}\cv{H}_{31}\v{v}_{1}\label{equ_condE5},\\
\v{v}_{4} &=\frac{\mu_{32}}{\kappa_{34}} \cv{H}_{34}^{-1}\cv{H}_{32}\v{v}_{2}\label{equ_condE6},
\end{align}
\begin{align}
\v{v}_{3} &=\frac{\mu_{41}}{\mu_{43}} \cv{H}_{43}^{-1}\cv{H}_{41}\v{v}_{1}\label{equ_condE7},\\
\v{v}_{3} &=\frac{\mu_{42}}{\kappa_{43}} \cv{H}_{43}^{-1}\cv{H}_{42}\v{v}_{2}\label{equ_condE8}.
\end{align}
\end{minipage}

Similar to the method introduced in Subsection \ref{subsection_Three_User_Network}, we are looking for ``loops'' in the above coupled system of linear equations. Inserting \eqref{equ_condE7}, \eqref{equ_condE4} and \eqref{equ_condE6} in \eqref{equ_condE1} as well as \eqref{equ_condE5}, \eqref{equ_condE3} and \eqref{equ_condE8} in \eqref{equ_condE2}, we can write
\begin{align}
\lambda_1 \v{v}_{2} &=\underbrace{\nu_{1} \cv{H}_{12}^{-1}\cv{H}_{13}\cv{H}_{43}^{-1}\cv{H}_{41}\cv{H}_{21}^{-1}\cv{H}_{24} \cv{H}_{34}^{-1}\cv{H}_{32}}_{\widehat{\cv{H}}_{\textrm{A}}}\v{v}_{2}\label{equ_condE9},\\
\lambda_2 \v{v}_{2} &=\underbrace{\nu_{2} \cv{H}_{12}^{-1}\cv{H}_{14}\cv{H}_{34}^{-1}\cv{H}_{31}\cv{H}_{21}^{-1}\cv{H}_{23} \cv{H}_{42}^{-1}\cv{H}_{43}}_{\widehat{\cv{H}}_{\textrm{B}}}\v{v}_{2}\label{equ_condE10},
\end{align}
with $\nu_{1}=\frac{\mu_{13}}{\mu_{12}}\frac{\mu_{41}}{\mu_{43}}\frac{\mu_{24}}{\kappa_{21}}\frac{\mu_{32}}{\kappa_{34}}\lambda_1$ and $\nu_{2}=\frac{\mu_{14}}{\kappa_{12}}\frac{\mu_{31}}{\mu_{34}}\frac{\mu_{23}}{\mu_{21}}\frac{\mu_{42}}{\kappa_{43}}\lambda_2$.
Because some of the channel matrices (e.g. $\cv{H}_{13}$) appear only in one of the matrices $\widehat{\cv{H}}_{\textrm{A}}$, $\widehat{\cv{H}}_{\textrm{B}}$ and because those matrices are random and independent, with probability 1, there is no solution that simultaneously fulfills \eqref{equ_condE9} and \eqref{equ_condE10} for any $\nu_1$ and $\nu_2$. Thus, we conclude that interference alignment with $d_i=1\,\forall i$ is not feasible in this case.

\section{Towards a General Achievability Criterion}
\label{section_Conjecture}

No general condition for the achievability of interference alignment in constant MIMO interference channels has been found so far. Using the iterative algorithm from \cite{Gomadamit08}, it appears that the existence of a solution depends solely on the dimensions of the problem ($K, \NT,$ $\NR, d_1\ldots d_K)$, and not on the particular channel realization.

Furthermore, for the particular case of $d_1=\ldots=d_K=1$, the numerical algorithm appears to converge (see Table \ref{table:AchievIA_d1})  iff
\begin{align}\label{equ_IA_conj}
\NR+\NT-1 \geq K.
\end{align}
This criterion is also consistent with a counting argument applied to the degrees of freedom of the system of non-linear equations in \eqref{equ_IAcondA}. We conjecture that eq. \eqref{equ_IA_conj} provides indeed an upper bound on the number of users $K$ that can achieve interference alignment with one degree of freedom each, over $\NR\times\NT$ constant channels. Further work on a proof of \eqref{equ_IA_conj} is ongoing.
\begin{table}[h]
\renewcommand{\arraystretch}{1.1}
\setlength{\doublerulesep}{1pt}
\begin{center}
\begin{tabular} {|c|cccccc|} \hline
\backslashbox{$N$}{$K$} & 3 & 4 & 5 & 6 & 7 & 8 \\
\hline
2 & \checkmark$^{*}$ &  &  &  &  &  \\
3 & \checkmark & \checkmark$^{*}$ & \checkmark &  &  &  \\
4 & \checkmark & \checkmark & \checkmark$^{*}$ & \checkmark & \checkmark &  \\
\vdots & &&&&&\\
\hline
\end{tabular}
\caption{Achievability ($\checkmark=$yes) of IA for $\NT=\NR=N$ and $d_i=1\,\forall i$.}
\label{table:AchievIA_d1}
\end{center}
\end{table}

Note that for the square channels ($\NT=\NR=N$) considered in Section \ref{subsection_Equivalent_Eigenvalue_Problem}, our constructive method only provides solutions for the case $K=N+1$ ($\checkmark^{*}$ in Table \ref{table:AchievIA_d1}), while \eqref{equ_IA_conj} suggests that a better scaling of the method is possible, since solutions can be found numerically iff $K\leq 2N-1$.

\section{Conclusion}
\label{section_conclusion}
We have introduced a constructive proof of the achievability of interference alignment, with one interference-free dimension per user for an arbitrary number of users $K$, for the interference channel with constant $N \times N$ MIMO channels, for the case where $N=K-1$. The closed-form solutions provide insight into the interference alignment problem. We also discussed the general feasibility of IA with one degree of freedom per user for various channel dimensions, and introduced a conjecture for a general feasibility criterion.





%


\begin{thebibliography}{1}

\bibitem{maddahali_isit06}
M.~Maddah-Ali, A.~Motahari, and A.~Khandani,
\newblock ``Signaling over {MIMO} multi-base systems: Combination of
  multi-access and broadcast schemes,''
\newblock in {\em Proc. IEEE Int. Symp. Information Theory (ISIT)}, Seattle,
  WA, USA, July 2006, pp. 2104--2108.

\bibitem{Jafarit08}
Syed~A. Jafar and Shlomo Shamai,
\newblock ``{D}egrees of {F}reedom {R}egion of the {MIMO X} {C}hannel,''
\newblock {\em IEEE Trans. Information Theory}, vol. 54, no. 1, pp. 151--170,
  Jan. 2008.

\bibitem{Cadambeit08}
Viveck~R. Cadambe and Syed~A. Jafar,
\newblock ``{I}nterference {A}lignment and {D}egrees of {F}reedom of the
  \textit{{K}}-{U}ser {I}nterference {C}hannel,''
\newblock {\em IEEE Trans. Information Theory}, vol. 54, no. 8, pp. 3425--3441,
  Aug. 2008.

\bibitem{Gou08}
Tiangao Gou and Syed~A. Jafar,
\newblock ``{D}egrees of {F}reedom of the \textit{{K}}-{U}ser \textit{{M}}
  {\textsf{x}} \textit{{N}} {MIMO} {I}nterference {C}hannel,''
\newblock {\em IEEE Trans. Information Theory}, 2008,
\newblock submitted.

\bibitem{cadambe-2009}
Viveck~R. Cadambe, Syed~A. Jafar, and Chenwei Wang,
\newblock ``Interference alignment with asymmetric complex signaling - settling
  the {Host-Madsen-Nosratinia} conjecture,'' 2009.

\bibitem{Gomadamit08}
Krishna Gomadam, Viveck~R. Cadambe, and Syed~A. Jafar,
\newblock ``{A}pproaching the {C}apacity of {W}ireless {N}etworks through
  {D}istributed {I}nterference {A}lignment,''
\newblock {\em IEEE Trans. Information Theory}, 2008,
\newblock submitted.

\end{thebibliography}

\end{document}